\documentclass[twocolumn,prl,aps,showpacs,superscriptaddress]{revtex4}
\usepackage{graphicx}
\usepackage{amsmath}

\newcommand{\mycol}{1}

\begin{document}

\title{Multiphoton coherent manipulation in large-spin qubits}

\author{S. Bertaina}
\email{sylvain.bertaina@im2np.fr}
\altaffiliation{Present address: IM2NP-CNRS (UMR 6242), 13397
Marseille Cedex, France.}
\affiliation{Physics Department and the National High Magnetic Field Laboratory, Florida State University, 1800 E. Paul Dirac Drive, Tallahassee, Florida 32310}

\author{L. Chen}
\affiliation{Physics Department and the National High Magnetic Field Laboratory, Florida State University, 1800 E. Paul Dirac Drive, Tallahassee, Florida 32310} 
\author{N. Groll}
\affiliation{Physics Department and the National High Magnetic Field Laboratory, Florida State University, 1800 E. Paul Dirac Drive, Tallahassee, Florida 32310} 
\author{J. Van Tol}
\affiliation{Physics Department and the National High Magnetic Field Laboratory, Florida State University, 1800 E. Paul Dirac Drive, Tallahassee, Florida 32310} 
\author{N. Dalal}
\affiliation{Physics Department and the National High Magnetic Field Laboratory, Florida State University, 1800 E. Paul Dirac Drive, Tallahassee, Florida 32310}
\affiliation{Department of Chemistry and Biochemistry, Florida State University, Tallahassee, FL 32306} 
\author{I. Chiorescu}
\email{ic@magnet.fsu.edu} 
\affiliation{Physics Department and the National High Magnetic Field Laboratory, Florida State University, 1800 E. Paul Dirac Drive, Tallahassee, Florida 32310}

\date{Submitted 10 October 2008, published 3 February 2009. Phys. Rev. Lett. \textbf{102}, 050501 (2009)}

\begin{abstract}
Large spin Mn$^{2+}$ ions ($S=5/2$) diluted in a non-magnetic MgO matrix of high crystalline symmetry are used to realize a six level system that can be operated by means of multi-photon coherent Rabi oscillations. This spin system has a very small anisotropy which can be tuned \textit{in-situ} to reversibly transform the system between harmonic and non-harmonic level configurations. Decoherence effects are strongly suppressed as a result of the quasi-isotropic electron interaction with the crystal field and with the $^{55}$Mn nuclear spins. These results suggest new ways of manipulating, reading and resetting spin quantum states which can be applied to encode a qubit across several quantum levels.
\end{abstract}

\pacs{03.67.-a,71.70.Ch,75.10.Dg,76.30.Da}

\maketitle

Quantum algorithms offer the possibility of solving certain problems with a tremendous increase of speed compared to their classical counterparts \cite{DiVincenzo2000}. Manipulating quantum bits, however, is a delicate process, and photons are an ideal choice as they interact with quantum systems in predictable ways. As qubit implementations, spin-based systems are nanoscopic, potentially operable up to room temperature\cite{Akimov2007}, and less noise sensitive and benefit of single-spin detection schemes\cite{Petta2005,Koppens2006,Atature2007}. When diluted enough to avoid spin-spin interactions, a variety of systems show long coherence times, \emph{e.g.}, the nitrogen-vacancy (N-V) color centers in diamonds\cite{Hanson2008,Dutt2007}, N atoms in C$_{60}$\cite{Morley2007}, Ho$^{3+}$ and Cr$^{5+}$ ions\cite{Bertaina2007,Nellutla2007} and molecular magnets\cite{Ardavan2007,Bertaina2008}. In this Letter, we present first observations of multi-photon spin coherent manipulation in a multi-level system (Mn$^{2+}$) diluted in MgO, a highly symmetric nonmagnetic matrix.

Such multi-state systems are proposed for quantum algorithms in either size-limited \cite{Leuenberger2001} or scalable \cite{Grace2006} schemes and to study new exotic quantum phenomena, like the quantum antiresonances \cite{Hicke2007}, by using microwave (MW) pulses to generate entangled states. The high symmetry of the host ensures that the Mn$^{2+}$ spins are in an almost isotropic environment, which permits reaching a regime where the MW excitation [last term of Hamiltonian (\ref{eq:1})] and the induced Rabi splittings are comparable to spin's anisotropy. Mn$^{2+}$ ions are diluted in a single crystal of MgO with cubic symmetry $F_{m\bar{3}m}$ (lattice constant 4.216 \r{A}). They replace Mg in a concentration estimated at 1 ppm. The spin Hamiltonian is given by (refs. \cite{Low1957,Smith1968}):
\begin{eqnarray}\label{eq:1}
    H&=&a/6\left[ {S_x^4+S_y^4+S_z^4-S(S+1)(3S^2-1)/5} \right]\\
&&+g\mu_B\vec{H}_0.\vec{S}-A\vec{S}.\vec{I}+g\mu_B\vec{h}_{mw}.\vec{S}
\cos(2\pi ft) \nonumber
\end{eqnarray}
where $S_{x,y,z}$ are the spin projection operators, $g=2.0025$ is the gyromagnetic ratio, $a =$ 55.7~MHz is the anisotropy constant, $A = 244$~MHz is the hyperfine constant of $^{55}$Mn ($I = 5/2$), $h_{mw}$ and $f$ represent the MW amplitude and frequency respectively, and $\vec{H}_0$ is the static field ($\vec{H}_0\perp \vec{h}_{mw}$). The $g$, $A$ and $a$ parameters were determined experimentally by continuous wave spectroscopy at room temperature and are in excellent agreement with literature values \cite{Low1957,Smith1968}. All Electron Paramagnetic Resonance (EPR) measurements are performed using a Bruker Elexsys 680 spectrometer at X-band ($\sim$9.6~GHz) in both continuous-wave (cw) and pulsed modes. 

\begin{figure}
\centering
\includegraphics[bb=0 0 300 207,width=\mycol\columnwidth]{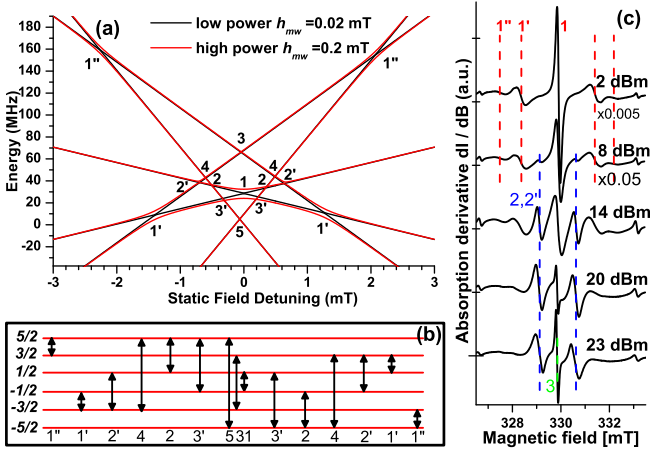}
\caption{(a) Dressed state energy diagrams for two $h_{mw}$ values, showing level repulsions (or Rabi splittings) when the Zeeman splitting matches the integer number of photons labeling each resonance. (b) Level diagram indicating the level spin projection (on the left) involved in each multi-photon transition. (c) CW absorption derivative at room temperature for increasing MW power (from 3 to 23~dBm; curves are shifted vertically). At low powers, only the central (1) and the laterally detuned single-photon resonances (1',1") are observed. At high powers, multi-photon resonances are revealed (2,2'-unresolved and 3 at the central resonance).}
\label{fig:1}
\end{figure}

Spin dynamics under coherent driving is studied using the rotating frame approximation \cite{Schweiger2001} (RFA), which consists of rotating the Hamiltonian's reference frame with a frequency $f$ around the direction of the static field $\vec{H}_0$. Experimental conditions ensures $a < A \ll g\mu_BH_0S/h = f = 9.62$~GHz, where $h$ is Planck's constant. This implies that (i) $\vec{H}_0$'s direction can be approximated as a good quantization axis for $S$ and $H$ and (ii) coherent MW driving is confined between levels of same nuclear spin projection $m_I$ (see also refs. \cite{Hicke2007,Schweiger2001}). A numerical diagonalization of the static Hamiltonian \cite{Stoll2006} is performed, and the truncation to the desired $m_I$ subspace leads to a diagonal Hamiltonian $H_{m_I}$. The later is used to obtain the RFA Hamiltonian $H_{rot}$ (see Eq. (30) in ref. \cite{Leuenberger2003}). The resulting time-dependent density matrix $\rho(t,T)$, where $T$ is temperature, gives the ensemble-averaged magnetization  $\langle Sz\rangle(t)=Tr\left(\rho(t,T) S_z\right)$. Also, the energy diagrams of the dressed states (spin-states in interaction with the microwave field \cite{Cohen-Tannoudji2004}) are computed by diagonalizing $H_{rot}$ as a function of the detuning  $\delta H=H_0-H_{res}$, where $H_{res}$ is the field of the central resonance $S_z=-1/2\leftrightarrow +1/2$ (\textit{e.g.}, $H_{res}=329.91$~mT; see Fig.~\ref{fig:1}).

The corresponding diagrams for two $h_{mw}$ values are given in Fig.~\ref{fig:1}a as a function of $\delta H$ for $\vec{H}_0 || [1 1 0]$ axis. At low MW powers, the dressed state diagram shows level crossings where resonances in a typical Zeeman diagram appear ($hf$ is subtracted from consecutive levels). Higher MW powers induce splittings of size $hF_R$, where $F_R$ is the Rabi frequency of the spin's coherent nutation. The participating number of photons is indicated near each resonance and the initial and final $S_z$ projections are shown in Fig.~\ref{fig:1}b. At zero detuning, the one-, three- and five-photon resonances are active with the 1-photon largely dominating the dynamics [see the size of the Rabi splittings in Fig.~\ref{fig:1}a]. At larger detunings, one observes two 1-photon resonances (1',1") and a 3-photon (3'), a two-, a four- and another two-photon (2') resonances of relatively low splittings. The existence of quasi-harmonic intermediate levels ensures, however, sizeable Rabi splittings even for high numbered resonances. Room temperature continuous wave (cw) spectroscopy indicates the existence of single-, double-, and triple-photon resonances when scanning the detuning of the static field as in Fig.~\ref{fig:1}c. The field scan is performed for the $m_I= -3/2$ group of levels. Similar peak structures exist for all other $m_I$ projections and are separated by $\sim$8.7~mT (see also ref. \cite{Sorokin1958}). At low MW power, the absorption derivative indicates a strong central (1) and two unresolved lateral (1' and 1") one-photon resonances. When increasing the nominal MW power from 3 to 23 dBm, the signal from the single-photon resonances decreases due to a saturation effect, while double- (2, 2') and triple-photon (3') spectroscopy peaks are revealed. The dashed lines indicate the resonances and are at symmetrical detunings as in Fig.~\ref{fig:1}a.

\begin{figure}
\centering
\includegraphics[width=\mycol\columnwidth]{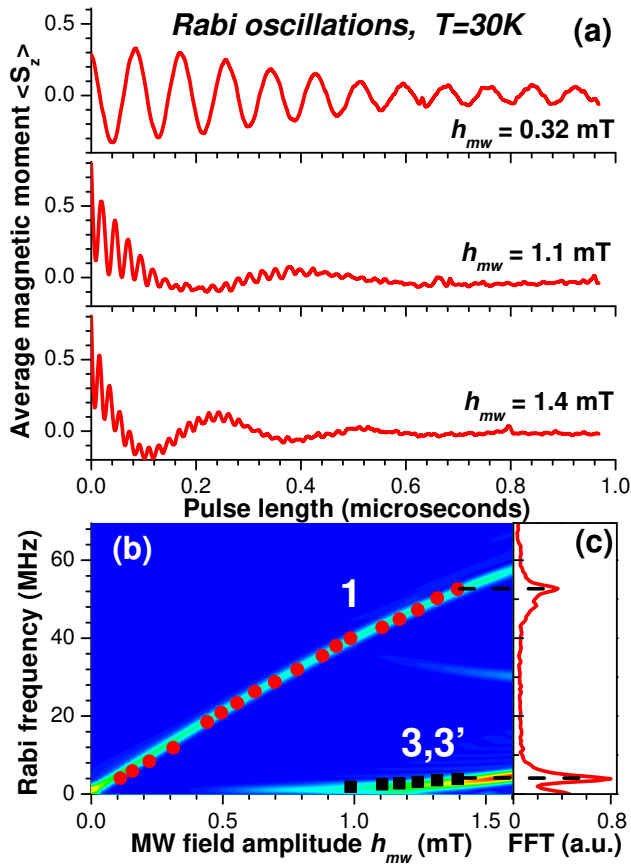}
\caption{(a) Rabi oscillations at $T$=30~K. At low power $h_{mw}=0.32$~mT, the one-photon oscillations show a decay time of $\sim$0.5~$\mu s$. At higher powers a slow oscillation is superposed on the one-photon one and is attributed to a coherent three-photon oscillation, in agreement with simulations. (b) FFT peaks of the experimental Rabi oscillations (red dots - single photon, black squares $-$ triple-photon oscillations). The full width at half maximum (FWHM) of the FFT peaks are the error bars, here similar to the symbol size. The contour plot shows FFT traces of simulated Rabi oscillations with $h_{mw}$ as single fit parameter. The color code goes from 0 (blue) to 1 (red) with mixed colors for intermediate values. (c) Experimental FFT trace for $h_{mw}=1.4$~mT showing two distinct peaks.}
\label{fig:2}
\end{figure}

Coherent Rabi driving \cite{Rabi1937} is achieved by setting the static field at the central resonance and applying one resonant MW pulse. After each pulse, the average magnetization $\langle S_z\rangle$ is measured by a spin-echo sequence \cite{Hahn1950,Bertaina2008}. This ensures an ensemble averaging over the active spin packets of the single crystal. $\langle S_z\rangle$ evolution as a function of time is shown in Fig.~\ref{fig:2}a for three values of $h_{mw}$ at $T=$30~K. The static field is applied along the unit vector $(1,0.8,1)/\sqrt 2.64$ and from here on data is taken for the $m_I=-3/2$ resonances. The Fast Fourier Transform (FFT) of the coherent motion indicates peaks for each Rabi splitting, as shown in Fig.~\ref{fig:2}b with red dots (one-photon) and black squares (three-photon driving). The contour plot represents simulated FFT, and an experimental FFT at the maximum power is shown in Fig.~\ref{fig:2}c. The simulations are in excellent agreement with the experimental findings. The MW field amplitude is the single fitting parameter giving the $h_{mw}$ values of Fig.~\ref{fig:2}a. At increased powers $h_{mw}$ =1.1~mT and 1.4~mT, the three-photon coherent driving is visible as a slow oscillation superposed on the fast single-photon one. The 3' Rabi splitting is slightly detuned but is visible due to the finite linewidth of the resonance. The mixing of several levels at high power adversely affects the Rabi decay time, but one notes, however, a large amplitude three-photon oscillation.
 
\begin{figure}
\centering
\includegraphics[width=\mycol\columnwidth]{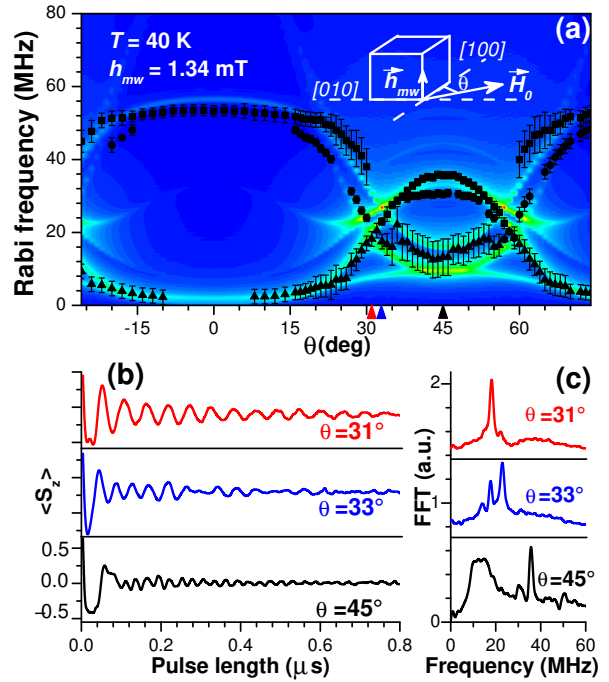}
\caption{(a) FFT peaks, in black, of Rabi oscillations recorded at $T$=40 K, $h_{mw}=1.34$~mT, and $m_I=-3/2$, as a function of $\theta$ (the inset shows the relative angles between the magnetic fields and the crystal axes). Error bars indicate peaks FWHM. The contour plot represents FFT traces of simulated Rabi oscillations (color code as in Fig.~\ref{fig:2}). The markers on the ordinate axis indicate the angles used in (b) and (c). (b) At the compensation angle $\theta=31^\circ$, the multi-level system behaves as a two-level system showing a single-valued Rabi frequency. At $\theta=33^\circ$ beatings between two close Rabi frequencies are observed. At $\theta=45^\circ$, $\langle Sz\rangle$ undergoes a fast nutation superposed on a slower and less defined one. (c) FFT of data in (b) [curves are shifted vertically, as in (b)].}
\label{fig:3}
\end{figure}
The anisotropy term $a$ acts as a small perturbation in the spin Hamiltonian but impacts significantly the spin dynamics. By means of crystal rotation, one can study the role of the anisotropy at fixed static and MW field amplitude. Rabi oscillations are recorded each 1$^\circ$ and analyzed by FFT. The angle dependence of Fig.~\ref{fig:3} presents experimental FFT peaks in the $m_I=-3/2$ case (symbols in black) superimposed on a calculated FFT contour plot with $h_{mw}$ as single fit parameter. The temperature is $T$=40~K, and the field direction is along the unit vector $k=(\cos\theta , \sin\theta , 0)$ [see the inset in Fig.~\ref{fig:3}a]. As expected from the cubic symmetry, the data shows a 90$^\circ$ periodicity in $\theta$. For $|\theta |\leq 20^\circ$, the one-photon oscillation (squares) dominates the dynamics at a rate $F_R^1 \sim $53.5 MHz corresponding to $h_{mw}$=1.34 mT. This identifies the dominant transition as being between the $S_z$ projections $-1/2\rightarrow+1/2$ [see resonance (1) in Fig.~\ref{fig:1}b] in agreement with the one-photon Rabi frequency given by (ref.~\cite{Schweiger2001}):
\begin{equation}\label{eq:2}
    hF_R^1=\frac{1}{2}g\mu_Bh_{mw}\sqrt{S(S+1)-S_z(S_z+1)}
\end{equation}
The relatively high value of the MW drive reveals three-photon oscillations for $\theta$ closer to 20$^\circ$ and above (the black triangles). Of particular interest is the region 25$^\circ<\theta <65^\circ$ containing two nodal points at $\theta = 31^\circ$ and 59$^\circ$ for which the effect of transverse anisotropy is compensated by field's direction\cite{Low1957}. At these compensation angles anisotropy effects are annihilated and the level system becomes harmonic. All levels with $S_z=-5/2$ to $+5/2$ are brought into a coherent superposition. Consequently, it behaves as an effective two-level system. The measured Rabi oscillation and its FFT at the compensation angle [see also the colored markers on the ordinate axis of Fig.~\ref{fig:3}a] are given in Fig.~\ref{fig:3}b and c respectively, showing a single mode oscillation with frequency $F_R^{comp}=$18.07~MHz. Note that $F_R^{comp}$ is given by the square root's pre-factor of Eq.~\ref{eq:2} which depends linearly on $h_{mw}$, in agreement with a two-level picture. The experimental ratio $F_R^1/F_R^{comp}$ = 2.96 is in agreement with an expected value of 3 [see Eq.\eqref{eq:2}]. The angular region where the anisotropy is compensated is found to be narrow, within 1$^\circ$. For angles slightly shifted away from the compensation angle, anisotropy generates a bi-modal Rabi oscillation, as the one presented in Fig.~\ref{fig:3}b for $\theta=33^\circ$, with a beating period given by modes difference. Accordingly, the corresponding FFT indicates two peaks located at 17.58 MHz and 22.95 MHz [Fig.~\ref{fig:3}c]. 

In the region between the two compensation angles $31^\circ <\theta < 59^\circ$, one observes numerically a complicated structure of multi-photon Rabi splittings developing in the dressed state energy diagrams. Experimentally, the Rabi oscillations do show a multi-mode structure, as the one measured for $\theta = 45^\circ$ [Fig.~\ref{fig:3}b]. Its FFT indicates a narrow peak at 35.7~MHz corresponding to the single photon drive and a broad peak centered at $\sim$13~MHz comprising three- and five-photon transitions [Fig.~\ref{fig:3}c].

\begin{figure}
\centering
\includegraphics[width=\mycol\columnwidth]{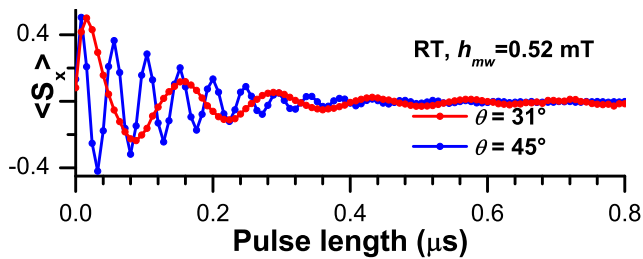}
\caption{Rabi oscillations measured at room temperature and $h_{mw}=$0.52~mT, for $\theta=31^\circ$ (in red) $-$ at the compensation angle, when the multi-level systems behaves as a two-level system $-$ and for $\theta=45^\circ$ ($\vec{H}_0||[101]$, in blue).  }
\label{fig:4}
\end{figure}
The phenomena described here are visible at room temperature (RT) as well, though with less intensitiy. As an example, Rabi oscillations at the compensation angle $\theta=31^\circ$ and also at $\theta=45^\circ$ are given in Fig.~\ref{fig:4} for the case $m_I=-1/2$. Here, the $\langle S_x\rangle$ projection is measured by the Free Induction Decay (FID) method immediately after the end of the driving MW pulse\cite{Nellutla2007}. As discussed above, the ratio of the Rabi frequencies (here 21~MHz/7~MHz) fulfills the condition $F_R^1/F_R^{comp}=3$.

Multi-photon oscillations are shorter lived when compared to the one-photon one due to the larger number of levels involved, as visible in Fig.~\ref{fig:2}. However, the slow oscillation attributed to the three-photon resonance remains coherent up to 0.4~$\mu s$. This shows that an important part of the noise spectrum is located in the high frequency range and affects nutations with $F_R>20$~MHz. Note that the large hyperfine field of the $^{55}$Mn has a very limited influence on the decoherence of the electronic spin owing to an isotropic hyperfine coupling \emph{A}. Similarly, the host's cubic symmetry and Mn$^{2+}$'s zero orbital momentum ensure a negligible spin-orbit coupling which decreases relaxation processes. Dipolar Mn$^{2+}$ $-$Mn$^{2+}$ couplings are estimated to be rather weak in this diluted system ($\sim$ 0.01 MHz) due to an average spin-spin distance of $\sim$42~nm. The main source of decoherence resides on the superhyperfine coupling between the Mn$^{2+}$ spins and the $^{25}$Mg nuclear spin bath\cite{Prokof'ev2000} ($I=5/2$ with 10\% natural abundance).

In this system level coupling and anharmonicity can be tuned by field orientation and MW power. For instance, consecutive MW pulses of different intensities will couple different levels via controlled multi-photon processes. Sweeping the field orientation through the compensation angle where levels are equally spaced could provide a means of resetting the qubits or additional gating capabilities, depending of the MW power. The use of large spins can simplify significantly the implementation of quantum algorithms since they provide a natural multi-level (or multi-qubit) scheme or can be used to encode qubits across several quantum levels. Coupling can be achieved by dipolar interactions at higher concentrations, by photons when integrated in microcavities, or by current oscillations if spin detection is done by electronic transport.  Large spin qubit readout can be done in various ways, for instance as in quantum dots\cite{Petta2005,Koppens2006}, optically\cite{Atature2007} or by using surface plasmons \cite{Akimov2007}. Such implementations adapted to the case of diluted spins in single crystals remain to be explored.

These first observations of spin states dressed by coherent photons show the leading role of a small anisotropy on the dynamics of a spin qubit. Our results apply to other spin qubit implementations with quasi-harmonic energy diagram and in quasi-isotropic environments. 

We thank S. Nellutla, J.S. Brooks and B. Barbara for fruitful discussions and M. Dykman (MSU) for valuable input. This work was supported by NSF Cooperative Agreement Grant No. DMR-0654118, NSF grants No. DMR-0645408, No. DMR-0506946, and No. DMR-0701462, the FSU and the NHMFL (IHRP-5059), DARPA (HR0011-07-1-0031) and the Sloan Foundation.

\bibliography{multiphotonrabi_Mn}

\end{document}